\documentclass[12pt,preprint]{aastex}

\newcommand{\liemail}{litp@tsinghua.edu.cn}
\newcommand{\liuemail}{liuhao@ihep.ac.cn}

\def \<{\langle}
\def \>{\rangle}

\shorttitle{Scan Induced CMB Anisotropy}
\shortauthors{Li \& Liu}

\begin{document}
\title{Observational Scan Induced Artificial CMB Anisotropy}
\author{Hao Liu\altaffilmark{1} and Ti-Pei Li\altaffilmark{1,2,3}}
\altaffiltext{1}{Key Laboratory of Particle Astrophysics, Institute of High Energy Physics, Chinese Academy of Sciences, Beijing, China; \liuemail}
\altaffiltext{2}{Department of Physics and Center for Astrophysics, Tsinghua University, Beijing, China; \liemail}
\altaffiltext{3}{Department of Engineering Physics and Center for Astrophysics, Tsinghua University, Beijing, China}

\begin{abstract}
To reliably detect the cosmic microwave background (CMB) anisotropy is of great importance in understanding the birth and evolution of the Universe. One of the difficulties in CMB experiments is the domination of  measured CMB anisotropy maps by the Doppler dipole moment from the motion of the antenna relative to the CMB. For each measured temperature the expected dipole component has to be  calculated separately and then subtracted from the data. A small error  in dipole direction, antenna pointing direction, sidelobe pickup contamination, and/or timing synchronism, can raise significant deviation in the dipole cleaned CMB temperature. After a full-sky observational scan, the accumulated deviations will be structured with a pattern closely correlated to the observation pattern with artificial anisotropies on large scales, including artificial quadrupole, octopole etc in the final CMB map.  Such scan-induced anisotropies on large scales can be predicted by the true dipole moment and observational scan scheme. Indeed, the expected scan-induced quadrupole pattern of the WMAP mission is perfectly in agreement with the published WMAP quadrupole. With the scan strategy of the Planck mission, we predict that scan-induced anisotropies will also produce an artificially  aligned quadrupole. The scan-induced anisotropy is a common problem for all sweep missions and, like the foreground emissions, has to be removed from observed maps. Without doing so, CMB maps from COBE, WMAP, and Planck as well, are not reliable for studying the CMB anisotropy.
\end{abstract}


\section{Introduction}
The cosmological principle states that the universe is homogeneous in the large-scale average. However, after the Wilkinson Microwave Anisotropy Probe (WMAP) first-year release, the 
quadrupole (with multipole moment $l=2$) and octopole ($l=3$) patterns of the cosmic microwave background (CMB) map were reported to be aligned along a particular axis \citep{teg04}. Later, with more significant statistics the quadrupole plane and the three octopole planes are far more aligned, three of these planes are orthogonal to the ecliptic, the normals to these planes are aligned with the direction of the dipole and with the equinoxes, and the remaining octopole plane is orthogonal to the supergalactic plane \citep{sch04}; the next two harmonics also appear to be aligned with the quadrupole and octopole \citep{lan05}. Such odd large-scale anomalies in CMB anisotropy, indicating the universe being arrayed around special axes, were also found by more works on the following released three-, five- and seven-year WMAP data.

For CMB experiments and for studies of the very early universe, it is greatly important to find out whether the curious behavior in the low $l$ CMB modes being just a coincidence produced by primordial fluctuations in our particular Hubble volume (by cosmic variance), or caused by a measurement error \citep{hin07,cho07}.
Here we point out that these "axes of evil" do exist due to the nature of CMB observations: the Doppler dipole is around the direction of the solar system's motion to the CMB; the inhomogeneous integration-time maps of both WMAP and Planck observations have characteristic patterns with the ecliptic plane being most sparsely observed and two poles over-sampled; furthermore, the dipole axis depends on the motions of the antenna to the solar system, the solar system to the Galaxy, and the Galaxy to the local supergalaxy, etc. Comparing with the CMB anisotropy, the Doppler dipole signal is very strong, thus its predicted amplitude has to be calculated and removed from the raw data of any observation during the scanning process, observation by observation. An error in the predicted dipole signal will cause a deviation in the dipole-cleaned temperature, which will be accumulated into artificial anisotropies in the final map after completing a full-sky survey.  These artificial anisotropies might carry traces of the dipole and scanning scheme; therefore, the combining effect of the dipole and scanning scheme has to be carefully inspected and removed in CMB experiments. On the other hand, the scan-induced anisotropy in CMB maps might also be useful in monitoring motions of our local universe.

In this paper, we show in \S2 that measurement and calculation errors relevant to the Doppler dipole can induce considerable deviations in observed CMB temperatures. We further demonstrate in \S3 that such dipole relevant deviations combining with observational scan pattern will produce artificial anisotropies in resultant CMB maps, particularly at the largest angular scales, and show how can such scan-induced pattern on large scales be predicted by the true dipole moment and observational scan scheme.  In \S4 we show that the expected scan-induced quadrupole pattern of the WMAP mission is perfectly in agreement with the published WMAP quadrupole. With the scan strategy of the Planck mission, we find that there will also exist an artificial aligned quadrupole. Finally, in \S5 we demonstrate that it's possible to remove the observational scan induced effect, and only after doing so can the CMB anisotropy be reliably  detected.

\section{Temperature Deviation from Doppler Dipole}
\label{sec:Temperature Deviation from Doppler Dipole}
\subsection{Dipole moment.}   For a sky pixel $p$ with the unit direction vector ${\bf n}(p)$, the Doppler signal
\begin{equation}
T_{_{d}}(p)=\frac{T_0}{c}\,{\bf v}\cdot{\bf n}(p)\,,
\label{equ:dipole moment}
\end{equation}
where $T_0=2.725$\,K is the monopole temperature, ${\bf v}$ the joint velocity of antenna relative to the CMB, $c$ the speed of light. The Doppler signal on all sky pixels gives a dipole moment with a pattern of half-hot, half-cold structure along a special axis, which strongly dominates observed full-sky microwave maps with amplitude greater than 3\,mK, two orders of magnitude  stronger than the CMB anisotropy.

\subsection{Dipole signal in measured temperature}
Let $p(t)$ denote the sky pixel pointed by the antenna at time $t$, the dipole component in the measured temperature can be calculated by
\begin{equation}
T_{_{d}}(p(t))=\frac{T_0}{c}\,{\bf v}(t)\cdot{\bf n}(p(t))=\frac{T_0}{c}\,{\bf v}^\prime(t)\cdot{\bf n}^\prime(p(t))\,,
\label{equ:dipole}
\end{equation}
where ${\bf v}(t)$ is the spacecraft velocity relative to the CMB rest frame, and ${\bf n}(p(t))$ is the unit direction, or line-of-sight (LOS), of the antenna at time $t$ in the CMB rest frame.  ${\bf v}^\prime(t)$ and ${\bf n}^\prime (p(t))$ are corresponding vectors in the spacecraft coordinate system.
To remove the dipole signal in CMB map, $T_d$ has to be evaluated for each observation by Eq.\,\ref{equ:dipole} and then be subtracted from the observed time-ordered data.

An LOS error will produce a pseudo signal in the calculated $T_d$
\begin{equation}
\Delta T_{_d}(p(t))=\frac{T_0}{c}\,{\bf v}(t)\cdot\Delta{\bf n}_{_L}(p(t))=\frac{T_0}{c}\,{\bf v}^\prime (t)\cdot\Delta{\bf n}_{_L}^\prime\,,
\label{equ:pseudo}
\end{equation}
where $\Delta{\bf n}_{_L}(p(t))$ and $\Delta{\bf n}_{_L}^\prime$ are the LOS error vectors in a sky coordinate system and in the spacecraft frame, respectively. In the spacecraft frame the LOS error $\Delta{\bf n}_{_L}^\prime$ is a constant vector. After subtracting the calculated dipole signal $T_{_{d}}(p(t))$, a temperature deviation $-\Delta T_{_d}(p(t))$ will be left in the resulting CMB temperature on the sky pixel $p(t)$.

Even if the real LOS is exactly determined, other observational errors can also produce pseudo dipole signal just like an LOS error. For example, we found a 25.6\,ms timing offset between the spacecraft attitude and radiometer output timestamps in the raw WMAP time-ordered data, which can generate a pseudo signal in calculating the Doppler dipole as if there were $\sim 7'$ LOS error (Liu, Xiong \& Li 2010). The error in the direction of CMB dipole, or in the vector $\bf{v}$ in Eq.\,\ref{equ:dipole}, can also produce a pseudo dipole signal like a LOS error. The CMB dipole determined by the COBE or WMAP mission has a direction error great than $\sim 6'$ \citep{ben03a}, which cannot be ignored in producing pseudo dipole signals.

\subsection{Sidelobe contamination from dipole.}
\label{sec:sidelobe}
A microwave antenna has response to a large solid angle, not merely in the main beam along the LOS. The complete sidelobe response can be described by a full sky map of normalized gain $G^\prime$ in the spacecraft coordinate system, in which the antenna is static and $G'$ does not change with time, with the summation of all gains (including the main beam) being equal to $N$ (the number of pixels in the map). The sidelobe pickup comes mainly from the dipole moment, which is the dominating signal source in the foreground-cleaned sky. For an observation to the sky pixel $p(t)$ at time $t$, the recorded signal is contaminated by the dipole induced sidelobe pickup
\begin{equation}
T_s(p(t))=\sum_{p'_s} G^\prime (p_s') T_d(p_s')/N\,,
\label{equ:pickup}
\end{equation}
where the summation goes over all sidelobe pixels $p_s'$ in the spacecraft frame, and $T_d(p_s')$ is the dipole amplitude of the sky pixel $p_s$. The sidelobe pickup $T_s(p(t))$ calculated by Eq.\,\ref{equ:pickup} also has to be subtracted from the observed temperature.  An error $\Delta G'$ in sidelobe response will produce a pseudo signal in the sidelobe pickup calculated by Eq.\,\ref{equ:pickup}
\begin{equation}
\Delta T_s(p(t))=\frac{T_0}{c}\,{\bf v'}(t)\cdot\sum_{p'_s}\frac{\Delta G'(p'_s)}{N}\,{\bf n'}(p'_s)=\frac{T_0}{c}\,{\bf v'}(t)\cdot \Delta {\bf n}'_s\,
\label{equ:pickup_error}
\end{equation}
where it's self-evident that $\Delta{\bf n}'_s$, the result of the summation, is a constant vector. Thus it can be seen as an equivalent LOS error. After subtracting the calculated sidelobe pickup, a temperature deviation $-\Delta T_s(p(t))$ will be left in the resulting CMB temperature at the sky pixel $p(t)$.

\subsection{Overall LOS error.}
In stead of Eq.~\ref{equ:pseudo} and Eq.~\ref{equ:pickup_error}, we can take $\Delta {\bf n}'=\Delta {\bf n}'_{_L}+\Delta {\bf n}'_s$ as an overall LOS error and use
\begin{equation}
\Delta T(p(t))=-\frac{T_0}{c}\,{\bf v}(t)\cdot\Delta{\bf n}(p(t))=-\frac{T_0}{c}\,{\bf v'}(t)\cdot \Delta {\bf n}'
\label{equ:error}
\end{equation}
to estimate the overall deviation in the observed temperature at pixel $p(t)$ after subtracting the dipole signal and sidelobe pickup contamination.

\section{Scan Induced Anisotropy}
\label{sec:scan}
In Eq.\,\ref{equ:error}, the velocity ${\bf v}(t)$ is determined  mainly by the solar system's movement with respect to the CMB, and is hence approximately a constant vector in the Galactic coordinate system. The overall LOS error $\Delta{\bf n}^\prime$  is a constant vector in the spacecraft frame. During a scan period that vectors $\Delta{\bf n}(p(t))$ are pointing within the hot-half/cold-half area of the dipole moment pattern in the Galactic coordinate system (in other words, the dot-product ${\bf v}^\prime (t)\cdot\Delta{\bf n}^\prime$ are positive/negative in the spacecraft frame),  temperature deviations $\Delta T(p(t))$ will be negative/positive and temperatures $T(p(t))$ lower/higher, and the corresponding sky pixels $p(t)$ will draw a continual cold/hot trace along the scan trajectory in the sky. Finally, when a full-sky survey is completed, all deviations $\Delta T(p(t))$ will be combined into a map. The final deviation $\Delta T(p)$ at a certain pixel $p$ can be calculated by
\begin{equation}
\Delta T(p)=\sum_{p(t)=p} \Delta T(p(t))/N_p
\label{equ:accumulation}
\end{equation}
with $N_p$ being the number of observations to the pixel $p$. The final deviation map is structured with a pattern that resembles the exposure map (see Fig.~\ref{fig:Planck deviation},~\ref{fig:wmap deviation}). 
Our IDL source code for calculating $\Delta T(p)$ has been made publicly available\footnote{http://dpc.aire.org.cn/data/wmap/09072731/release\_v1/planck\_sim/}.

 For a continuous-sweep mission, the design of scanning strategy is constrained by a number of technical factors, e.g. solar irradiation, on-board fuel consumption, contact with ground antennas etc., it is very difficult to achieve a homogeneous survey.  With inhomogeneous coverage, continuous-scan induced anisotropies should consequentially be inhomogeneous with artificial spherical harmonics of low-order moments. In fact, the WMAP team already realized that the large-scale non-Gaussian anisotropy features are very similar to the WMAP scan pattern \citep{spe06}, just as expected by the scan-induced anisotropy analyzed above, and we found that the pixel temperature is significantly correlated with the number of observations \citep{litp09}.

\section{The WMAP and Planck Missions}
\label{sec:missions}
\subsection{Planck}
\label{sec:planck mission}
\begin{figure}[t]
\includegraphics[width=1.0\textwidth]{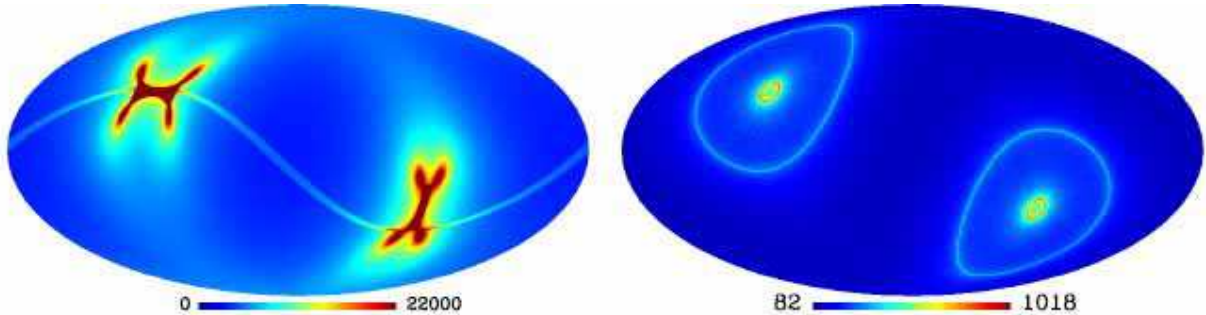}
\caption{Number of observations per pixel in Galactic coordinates. Simulated one-year scan for the Planck mission. {\it Right panel:} From the Q1-band one-year scan scheme of the WMAP mission.}
\label{fig:nobs}
\end{figure}

\begin{figure}[t]
\includegraphics[width=1.0\textwidth]{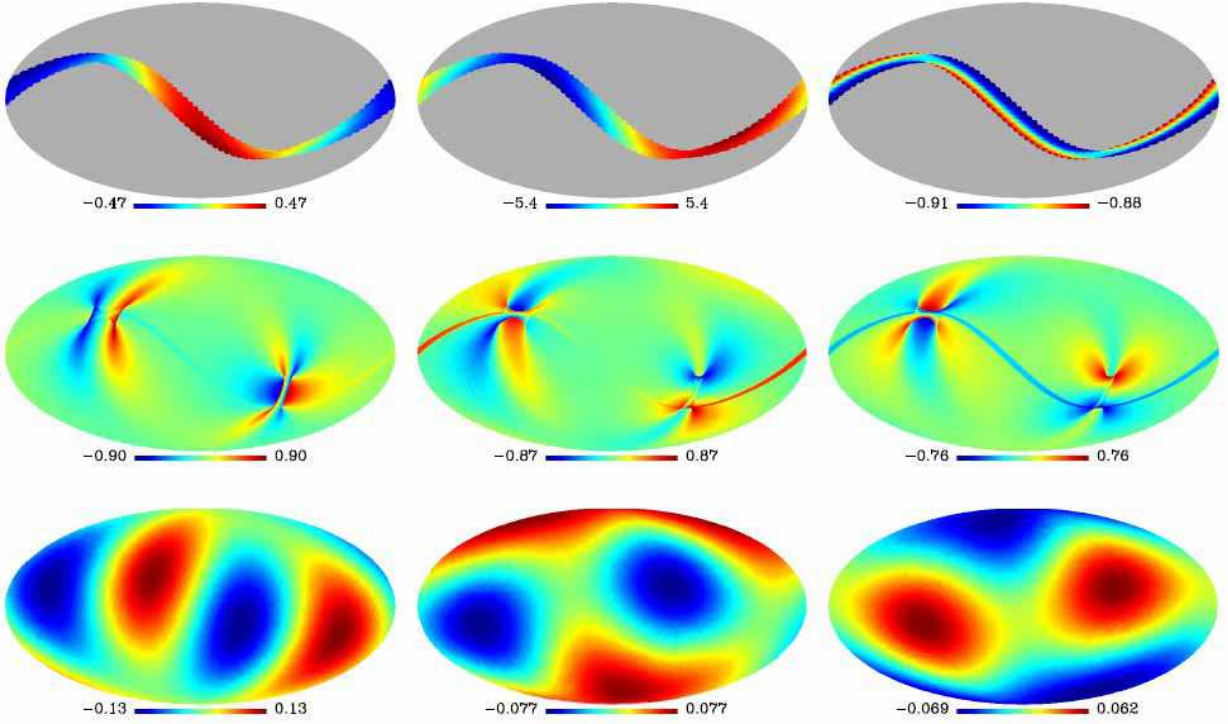}
\caption{Expected scan-induced deviation maps for the Planck mission with $1'$ overall LOS error.   From left to right: the overall LOS error is along the X, Y, Z-axis of the spacecraft coordinate frame, respectively. {\it Top panels:} Temperature deviation traces along simulated Planck 10-day scan path. {\it Middle panels:} Temperature deviations after a full-sky survey. {\it Bottom panels:} Quadrupole components of the upper panels. All in Galactic coordinates and the units are  $\rm{\mu k}$.}
\label{fig:Planck deviation}
\end{figure}

We produce a scan scheme $\{p(t)\}$ with the Planck scanning strategy \citep{Dup05} and HEALPix \citep{gor05} resolution parameter $N_{side}=1024$.  The left panel of Fig.~\ref{fig:nobs} shows the exposure map for a one-year full-sky survey.  Along with the simulated scan scheme, we calculate the temperature deviation series $\{\Delta T(p(t))\}$ by Eq.\,\ref{equ:error} with an assumed overall LOS error $\Delta {\bf n'}$ of $1'$ and along the X-, Y-, and Z-axis in the spacecraft coordinate system separately\footnote{In the spacecraft coordinates, the $X$-axis is parallel to plane of radiators, the $Z$-axis is the anti-sun direction of the spin axis, and the $Y$-axis is perpendicular to both.}. To visually show the trace formation process along with the Planck survey, the top panels of Fig.~\ref{fig:Planck deviation} present the traces $\{\Delta T(p(t))\}$ for a 10-day scanning, where we can see temperatures along a scan trajectory being really structured with alternately appearing hot and cold. All temperature deviations after one-year full-sky survey are combined by Eq.\,\ref{equ:accumulation} into deviation maps $\Delta T_{x}$, $\Delta T_{y}$, and $\Delta T_{z}$ shown in the middle panels of Fig.\,\ref{fig:Planck deviation}, where we can see the scan-induced deviations closely resembling the Planck exposure time map.

\begin{figure}[t]
\includegraphics[width=1.0\textwidth]{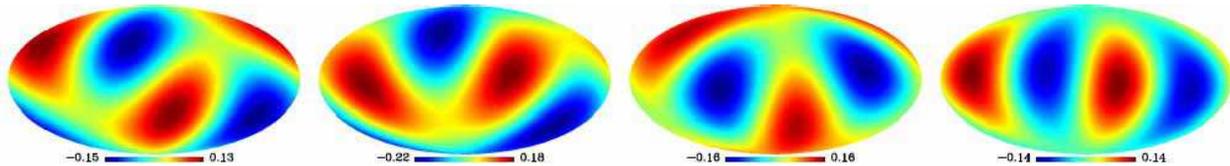}
\caption{Possible Planck quadruples}
\label{fig:expectation}
\end{figure}

The bottom panels of Fig.\,\ref{fig:Planck deviation} show the quadrupole components, $\Delta T_{q,x}$, $\Delta T_{q,y}$, and $\Delta T_{q,z}$,  of the three deviation maps shown in the middle panels. With these maps, the artificial quadrupole from any overall LOS error 
$\Delta {\bf n'}=(\delta_x,\delta_y,\delta_z)$ can be calculated by $\Delta T_q=\delta_x \Delta T_{q,x}+\delta_y \Delta T_{q,y}+\delta_z \Delta T_{q,z}$. Fig.\,\ref{fig:expectation} shows four possible Planck quadrupole patterns with $ (\delta_x, \delta_y, \delta_z)=(1,1,1)$, (1,1,-1), (1,-1,1), and (1,-1,-1), respectively.  It is expected from Fig.\,\ref{fig:expectation} that, due to scan-induced anisotropies, there may also exist an artificial quadrupole in Planck. It may also be aligned with an axis lying in the ecliptic, (like the case shown in the left panel of Fig.\,\ref{fig:expectation}), or in the Galaxy's plane (the right panel of Fig.\,\ref{fig:expectation}), or more possibly between the ecliptic and Galaxy's plane. Note that the exact templates of scan-induced anisotropies can be calculated when the real Planck data release is available. Thus the amplitudes shown here are only qualitative. The real Planck quadrupole depends on the real overall LOS error vector. Our calculations are based on the simulated Planck scanning scheme with assumed initial conditions of scan geometry. The real pattern of the scan-induced Planck quadrupole could hence be some kind of rotation or sign reversion to the ones shown here.

\subsection{WMAP}
\begin{figure}[t]
\includegraphics[width=1.0\textwidth]{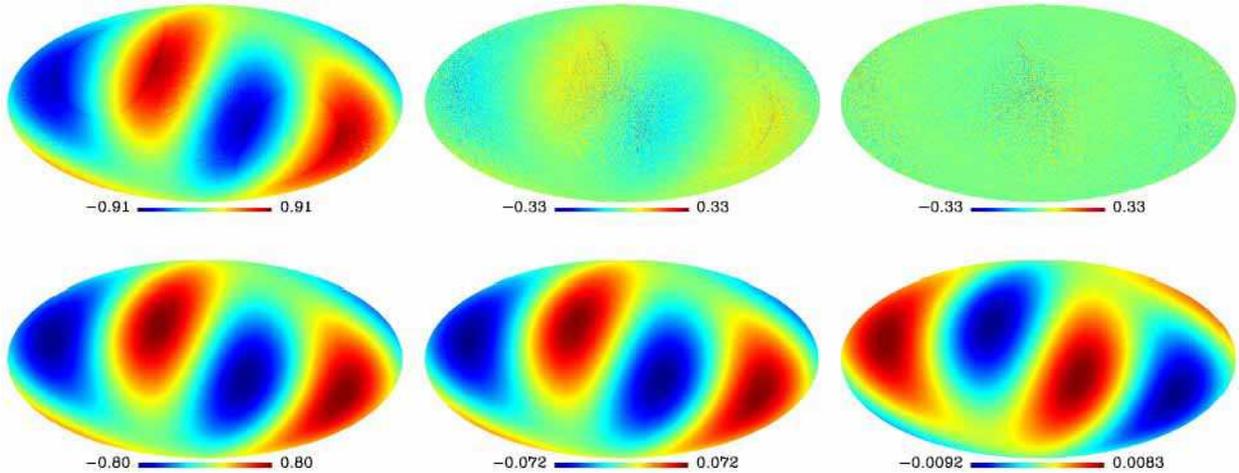}
\caption {Predicted scan-induced deviation maps for the WMAP mission with $1'$ overall LOS error using Q1-band one-year real scan scheme. From left to right: the overall LOS error is along the X, Y, Z-axis of the spacecraft coordinate frame, respectively. {\it Upper panels:} Temperature deviations. {\it Lower panels:} Quadrupole components of the upper panels. All in Galactic coordinates and the units are  $\rm{\mu k}$.}
\label{fig:wmap deviation}
\end{figure}

We use the WMAP Q1-band one-year scan scheme without any mask (because we don't use the temperature data)  to produce an expected scan-induced deviation map with the resolution parameter $N_{side}=512$, the resultant exposure map is shown in the right panel of Fig.\,\ref{fig:nobs}. Similar to the Planck mission, we calculate temperature deviation maps $\Delta T_x$, $\Delta T_y$ and $\Delta T_z$ shown in the upper panels of Fig.\,\ref{fig:wmap deviation}, where the bottom panels show their quadrupole components\footnote{The WMAP mission makes measurements with two antennas and records the difference between the two antenna temperatures. In this work, we have used two ways to estimate the temperature deviations relevant to the dipole issues in the WMAP mission: one by using the same simulation program for Planck (presented in footnote 1), and the other by using the real Q1-band one-year scan scheme (ignoring the mask). These two ways give consistent results, and the results shown here in Fig.\ref{fig:nobs} and Fig.\ref{fig:wmap deviation} are by the real scan scheme.}. Different to the Planck mission, the deviation components in Y- and Z-axis for the WMAP mission have much smaller amplitudes and the three quadrupole components are very similar to each other, thus we can predict the scan-induced quadrupole pattern more determinately. The prediction is really in agreement with the released WMAP quadrupole.

In \S~\ref{sec:Temperature Deviation from Doppler Dipole} and \S~\ref{sec:scan}, we illustrated that the LOS error, timing offset and sidelobe uncertainties can all cause temperature deviations. We find that an overall LOS error with amplitude of about $7'$ can produce most observed WMAP quadrupole (Liu, Xiong \& Li 2010). In previous works by  Roukema (2010) and Liu, Xiong \& Li (2011), two different methods have been independently used to detect possible timing offset effects in the temperature map and raw data    respectively, and fully consistent results that support the existence of such effects have been obtained.
On the other hand, a sidelobe uncertainty of about $10\%$ can also produce a similar artificial quadrupole alone. They can take effect simultaneously, and it's hard to exactly distinguish them, thus the best way to remove them together is probably by using a template-based approach (see \S5.2).

 As for the sidelobe uncertainty issue, a possible way is to improve the sidelobe measurement accuracy to see if the corresponding temperature deviation can be reduced. However, we have found that, although the WMAP team have provided sidelobe response data files in each official release, there is no essential improvement except for some minor posterior processing adjustment from WMAP1 to WMAP7. We have even seen that the K1-band sidelobe response files are exactly the same for all releases. Since the WMAP spacecraft has now stopped working, we probably need to wait for Planck for an improved and independent sidelobe measurement, so as to run a possible test on the sidelobe uncertainty issues.

It has been found that the WMAP CMB temperatures are significantly correlated with the corresponding number of observations \citep{litp09}. The scan induced artificial anisotropy is probably the most likely explanation to this phenomenon. With the same program used to detect the observation number correlation, we have checked the average absolute correlation strength between the simulated scan induced artificial anisotropy and the corresponding exposure map, which are 0.58 and 0.53 for WMAP and Planck respectively, strong enough to be able to leave some detectable scan-pattern like trace amongst real CMB, noise and foreground. Therefore, it is possible to check the scan induced artificial anisotropy by the correlation method in both missions. If the Planck image quality is really very good, then it will be even possible to see the characteristic pattern of scan induced artificial anisotropy directly in its result.

\section{Can CMB Anisotropy be Reliably Detected?}
\subsection{Can the overall LOS error be eliminated a priori?}
The equivalent LOS error described in \S\ref{sec:sidelobe} is a common problem in CMB experiments, because the Doppler dipole moment dominates all observed full-sky CMB maps. The only way to eliminate the equivalent LOS error a priori is to know the antenna 4$\pi$ response exactly, but this is very difficult: If the uncertainty of sidelobe response is about $100\%$, i.e. $\Delta G'(p_s')\sim G'(p_s')$, then with the data file of WMAP sidelobe response\footnote{The data file of WMAP sidelobe response can be found at http://lambda.gsfc.nasa.gov/product/map/dr4/farsidelobe\_info.cfm.}, the amplitude $\hat{\delta}$ of the equivalent LOS error  induced by sidelobe Doppler pickup (ignore the main lobe uncertainty) can be estimated by
\begin{equation} \hat{\delta}\simeq \arcsin(|\Delta{\bf n'}|/|{\bf n'}|)=\arcsin (|\sum_{p'_s}\frac{G'(p_s')}{N}\,{\bf n'}(p_s')|/|{\bf n'}|)\,,
\label{equ:los}
\end{equation}
which are $\sim 50' - 75'$ for different bands. From Fig.\,\ref{fig:wmap deviation} we see that $1'$ equivalent LOS error (requiring sidelobe error $<2\%$) is enough to produce $\sim 1\,\mu$K temperature deviation. However, this is for the sidelobe only. The main lobe response is much stronger than sidelobe: If we take the main lobe efficiency as $98\%$, then the overall main lobe response is about 50 times stronger than the overall sidelobe, and hence $<0.04\%$ main lobe response error can probably produce  temperature deviation with similar amplitude to $2\%$ sidelobe uncertainty. If we believe that the temperature deviation should be less than 1 $\rm{\mu K}$ for a highly reliable CMB estimation, then these two percentage data ($2\%$ for sidelobe and $0.04\%$ for main lobe) roughly describe the required antenna response accuracy.

For the sidelobe gain map of the first year WMAP observation, the overall calibration uncertainty is estimated as large as $\sim 30\%$ \citep{barnes03}.\footnote{This $30\%$ is actually an average level uncertainty, which means the total sidelobe efficiency is $30\%$ uncertain. Apparently, this is only part of the overall uncertainty which is relevant to not only the total efficiency, but also response on each pixel $p_s'$.}  Although we believe that the Planck mission has a much better performance, a requirement like $<2\%$ sidelobe uncertainty or $<0.04\%$ main lobe gain uncertainty might still exceed the equipment limitation. Besides the sidelobe contamination, $\sim 6'$ error in the Doppler dipole direction \citep{ben03a} also cannot be ignored.

Therefore, to get a reliable CMB map, the overall LOS error has to be considered and the scan-induced artificial anisotropy has to be removed posterior. Without doing so, CMB maps from sweep missions, including COBE, WMAP, and Planck as well, all can only be preliminary and yield unreliable results.

\subsection{Can scan-induced anisotropy be removed posterior?}
To remove the scan-induced anisotropy, the main difficulty is that the overall LOS error is unknown. In other words, the error vector $\Delta {\bf n'}=\delta_x\, \Delta{\bf n}_x'+\delta_y\, \Delta{\bf n}_y'+\delta_z\, \Delta{\bf n}_z'$ has three unknown coefficients $\delta_x$, $\delta_y$, and $\delta_z$. Is it possible to remove unfavorable signals with unknown intensities? The answer is yes. Several techniques can do it. For example, observed sky microwave maps are contaminated by diffusion foreground emission generated by different diffusion processes with known physical mechanisms but unknown intensities. The WMAP team used template fits with the unknown intensities as fitted parameters to remove these foreground emission effectively, and also used the ILC method -- using the internal linear combination of observed maps at different frequencies with estimated weights to suppress foreground and noise as far as possible \citep{ben03b}. Similar methods can also be used to diagnose the pseudo dipole and minimize the scan-induced anisotropy \citep{liu10,rou10}.

\begin{figure}[t]
\includegraphics[width=1.0\textwidth]{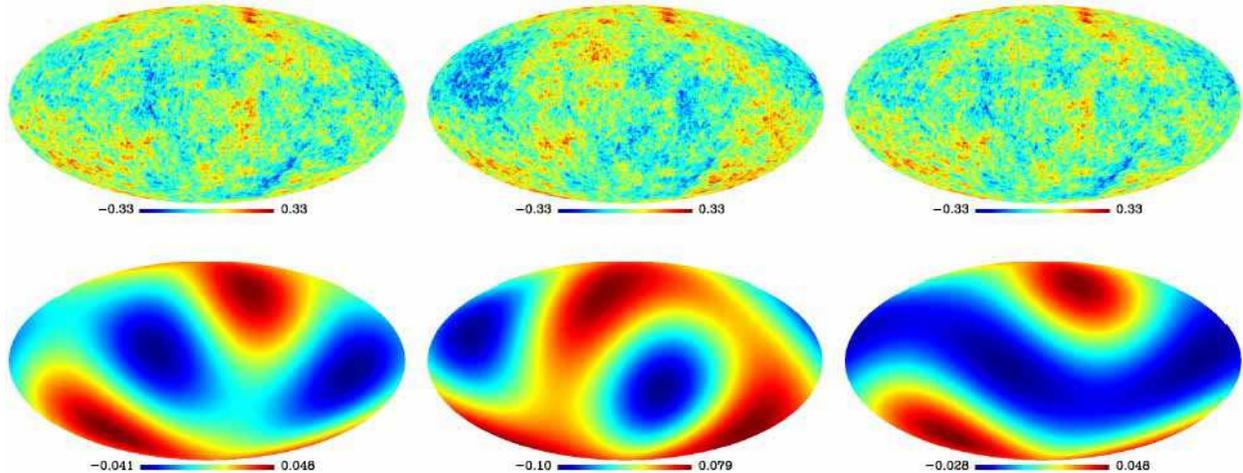}
\caption{Removing scan-induced anisotropy by template fits.   {\it Left panels:} Simulated $\Lambda$CDM model CMB map. {\it Middle panels:} Simulated observed map ($\Lambda$CDM model map + artificial anisotropies with WMAP scan scheme. {\it Right panels:} After template removal. {\it Upper panels:} Temperature map. {\it Lower panels:} Quadrupole component. All in Galactic coordinates and the units of temperature are mK.}
\label{fig:removing}
\end{figure}

For a CMB mission, we produce template maps $\Delta T_x$, $\Delta T_y$ and $\Delta T_z$ of scan-induced anisotropy from its scan scheme with an assumed value of $\delta$ and letting $\delta_x=\delta_y=\delta_z=\delta$. From the observed CMB map $T^*$ and with the templates maps, we can derive the cleaned map
\begin{equation}\label{equ:clean}
T=T^*-(c_x\, \Delta T_x+c_y\, \Delta T_y+c_z\, \Delta T_z)\,,
\end{equation}
where the coefficients $c_x$, $c_y$ and $c_z$ are determined by minimizing the variance of $T$. The result of template-based removal is not dependent on the amplitude $\delta$ that is assumed for the overall LOS  error. A larger amplitude leads to a lower fitted coefficient in template fitting so that the final correction to the CMB map is the same. To show the ability of scan-induced anisotropy removal by template fits, we simulate an observed CMB temperature map with the "synfast" routine in the HEALPix software package (available at http://healpix.jpl.nasa.gov) and a $\Lambda$CDM model power spectrum with a quadrupole of 1351\,$\mu$K$^2$ shown in the left panels of Fig.\,\ref{fig:removing}. To the temperature map we add deviations from the WMAP scanning scheme with an artificial quadrupole of 6492\,$\mu$K$^2$ to produce a simulated observed map, shown in the upper middle panel of Fig.\,\ref{fig:removing}. The lower middle panel shows the quadrupole component of the simulated observed map with a measured quadrupole of 5768\,$\mu$K$^2$. After template removal,  we get a cleaned map (the upper right panel of Fig.\,\ref{fig:removing}) and its quadrupole component with an amplitude of 1186\,$\mu$K$^2$ (the lower right panel of Fig.\,\ref{fig:removing}). We can see from Fig.\,\ref{fig:removing} that the template removal approach can effectively remove the scan-induced artificial effect from an observed map, and in both amplitude and structure pattern, the quadrupole component in the cleaned map is more reliable than that without the template removing, even the artificial quadrupole (6492\,$\mu$K$^2$) is much stronger than the real signal (1351\,$\mu$K$^2$).

\subsection{Can largest-scale anisotropies be predicted?}
For a CMB scan mission, template maps representing the distribution of scan-induced anisotropies can be produced with the observation scan scheme. Should the artificial anisotropy be removed with the templates before releasing a final CMB map?  The answer obviously should be yes. Unfortunately, in the meantime,  many would prefer to accept the odd "axis of evil" rather than a cleaned result after necessary correction. They worry about the corrected quadrupole being too low  because  the released WMAP quadrupole is already lower than expected from the standard cosmology model. In  exploring nature, however, to make experimental data reliable should be more important than to make it satisfy a theoretical requirement. Furthermore, in fact current inflation theories cannot predict the CMB fluctuation amplitude at the largest angular scale if the time when the scale left the horizon was as early as before the reheating period while the primordial density perturbations had not occurred. If it is true, it is just beyond, not against the standard model. Therefore, there is no a priori reason to reject a low fluctuation power at the largest scales for the sake of the standard model. For an observed full-sky CMB map, scan-induced artificial anisotropies should be minimized with predicted templates. More reliable CMB large-scale anisotropies cleaned from scan-induced deviations might be a new tool to study the very early universe, including the circumstances of the early epoch of inflation before the vacuum phase transition, and the start and evolution of the  primordial density fluctuations.

\acknowledgments
Prof. Charling Tao and the anonymous referee are thanked for helpful comments on the manuscript. This work is supported by the National Natural Science Foundation of China (Grant No. 11033003), the National Basic Research Program of China (Grant No. 2009CB824800). The data analysis made use of the WMAP data release (http://lambda.gsfc.nasa.gov/product/map/) and the HEALPix software package (http://healpix.jpl.nasa.gov).

\clearpage

\begin{thebibliography}{}
\bibitem[Barnes et al. 2003]{barnes03} Barnes, C., et al.\ 2003,  \apjs, 148, 51
\bibitem[Bennett et al. 2003a]{ben03a} Bennett, C. L. et al.\ 2003a, \apjs, 148, 1
\bibitem[Bennett et al. 2003b]{ben03b} Bennett, C. L. et al.\ 2003b, \apjs, 148, 97
\bibitem[Cho 2007]{cho07}Cho, A.\ 2007,  Science, 317, 1848
\bibitem[de Oliveira-Costa et al. 2004]{teg04}de Oliveira-Costa, A., Tegmark, M., Zaldarriaga, M. \& Hamilton, A.\ 2004,  Phys. Rev. D, 69, 063516
\bibitem[Dupac \& Tauber 2005]{Dup05} Dupac, X., \& Tauber, J.\ 2005,  \aap, 430, 363
\bibitem[Gorski et al. 2005]{gor05} Gorski, K. M. et al.,\ 2005, \apj, 622, 759
\bibitem[Hinshaw et al. 2007]{hin07} Hinshaw, G. et al.,\ 2007, \apjs, 170, 288
\bibitem[Land \& Magueijo 2005]{lan05}Land, K. \& Magueijo J.\ 2005, Phys. Rev. Lett., 95, 071301
\bibitem[Li et al. 2009]{litp09}Li, T.P., Liu H., Song L.M., Xiong S.L., \& Nie J.Y.\ 2009,  \mnras, 398, 47
\bibitem[Liu \& Li 2010]{liu10} Liu, H. \& Li, T. P.\ 2010,  Chinese Sci Bull, 55, 1; arXiv:1005.2352
\bibitem[Liu et al. 2010a]{LXL10}Liu H., Xiong S.L., \& Li T.P.\ 2010,  arXiv:1003.1073
\bibitem[Liu et al. 2010b]{lxl11} Liu, H., Xiong, S.L., \& Li, T.P.\ 2011, \mnras~Lett., in press
\bibitem[Roukema 2010]{rou10} Roukema, B. F.\ 2010, arxiv:1007.5307
\bibitem[Schwartz et al. 2004]{sch04} Schwartz, D., Starkeman, G., Huterer, D. \& Copi, C.\ 2004, Phys. Rev. Lett. 93, 221301
\bibitem[Spergel et al. 2006]{spe06} Spergel D. N. et al.\ 2006,  arXiv:astro-ph/0603449v1



\end{thebibliography}
\end{document}